\documentclass[aps,10pt,superscriptaddress,amsmath,amssymb,twocolumn,showpacs,floatfix,reprint]{revtex4-2}

\usepackage{braket}
\usepackage{subfigure}
\usepackage[latin9]{inputenc}
\usepackage{color}
\usepackage{verbatim}
\usepackage{amsthm,amsmath,amssymb}
\usepackage{mathrsfs}
\usepackage{graphicx}
\usepackage{esint}
\usepackage{mathrsfs}
\usepackage{bm}
\usepackage{hyperref}
\usepackage{mathrsfs}
\usepackage{times}

\definecolor{OliveGreen}{cmyk}{0.64, 0, 0.95, 0.40}
\definecolor{purple}{rgb}{0.6,0,0.5}

\begin{document}
\title{Chiral spin liquids with projected Gaussian fermionic entangled pair states}

\author{Sen Niu}\email{sen.niu@irsamc.ups-tlse.fr}
\affiliation{Laboratoire de Physique Th\'eorique, C.N.R.S. and Universit\'e de Toulouse, 31062 Toulouse, France}

\author{Jheng-Wei Li}
\affiliation{Universit\'e Grenoble Alpes, CEA, Grenoble INP, IRIG, Pheliqs, F-38000 Grenoble, France}

\author{Ji-Yao Chen}\email{chenjiy3@mail.sysu.edu.cn}
\affiliation{Guangdong Provincial Key Laboratory of Magnetoelectric Physics and Devices, Center for Neutron Science and Technology, School of Physics, Sun Yat-sen University, Guangzhou 510275, China}

\author{Didier Poilblanc}\email{didier.poilblanc@irsamc.ups-tlse.fr}
\affiliation{Laboratoire de Physique Th\'eorique, C.N.R.S. and Universit\'e de Toulouse, 31062 Toulouse, France}

\begin{abstract}

We study the parton construction of chiral spin liquids (CSLs) using projected Gaussian fermionic entangled pair states (GfPEPSs).
First, we show that GfPEPSs can represent generic spinless Chern insulators faithfully with finite bond dimensions.
Then, by applying the Gutzwiller projection to a bi-layer GfPEPSs, spin-1/2 Abelian and non-Abelian CSLs are obtained for Chern number $C=1$ and $C=2$, respectively. As a consequence of the topological obstruction for GfPEPSs, very weak Gossamer tails are observed in the correlation functions of the fermionic projected entangled pair state (PEPS) ansatze, suggesting that the no-go theorem for chiral PEPS is universal but does not bring any practical limitation. Remarkably, 
without fine tuning,
all topological sectors can be constructed showing the expected number of chiral branches in the respective entanglement spectra, providing a sharp improvement with respect to the known bosonic PEPS approach.
\end{abstract}

\date{\today }

\maketitle

\emph{Introduction.---}The notion of topological phase has revolutionized our understanding of phase of matter beyond the Landau paradigm. In two-dimensional systems without time-reversal symmetry, if there exists chiral edge modes moving only in one direction, the states are dubbed as chiral topological states. The most well-known chiral state in lattice free fermion systems is the Chern insulator \cite{haldane1988model,Qi2011RMP}, where the topology is completely characterized by the bulk Chern number $C$ indicating the number of chiral edge modes \cite{thouless1982quantized,hatsugai1993chern}. Through Gutzwiller projection on copies of Chern insulators (labeled by a spin index), a 
chiral spin liquid (CSL) state in the parton representation \cite{wen2002quantum} can be obtained. Interestingly, in contrast to their parent chiral Chern insulators, CSLs inherit long-range topological order 
from the Gutzwiller projection.
Hence, CSLs are bosonic variants of the fractional Quantum Hall states, and can be classified by the chiral gapless modes on the edge or, equivalently, the entanglement spectrum (ES)~\cite{li2008entanglement,qi2012general} described by (1 + 1)-dimensional Wess-Zumino-Witten (WZW) conformal field theories (CFT)~\cite{gawedzki1990}. 
For example, for two copies of half-filled Chern insulators with $C=1$ in each copy, the projected spin state becomes the topological $\mathrm{SU}(2)_1$ CSL \cite{kalmeyer1987,zhang2011topological} which is equivalent to the bosonic $\nu=1/2$ Laughlin wavefunction \cite{laughlin1983anomalous}. For more general cases, the topological nature of parton wave functions built from Chern insulators with higher Chern number are not clear, thus numerical methods 
for characterizing parton wavefunctions are desired.

The projected entangled pair states (PEPS) \cite{verstraete2004renormalization} have been successfully used for investigating two-dimensional topological states, where 
non-chiral topological orders can be encoded by gauge symmetry exactly~\cite{toriccode2009,schuch2012,chen2018,cirac2021}.  
However, there seems to exist a topological obstruction for PEPS to represent chiral topological states. 
For the case of free fermions where the corresponding PEPS representation is the Gaussian fermionic PEPS (GfPEPS), the obstruction has been proven exactly \cite{wahl2013projected,wahl2014symmetries,dubail2015tensor}, namely, if a GfPEPS is chiral then its bulk should be gapless. 
For the non-Gaussian case such as those in spin systems, a series of numerical studies show that the numerically optimized chiral bosonic PEPS also have artificial (gossamer) long-range correlations in the bulk~\cite{poilblanc2017,chen2018non,niu2022chiral,hasik2022}.
Since interacting chiral PEPS are also likely to be subject to topological obstruction, it becomes important to 
scrutinize the possible artifacts of the PEPS descriptions of a true CSL. In particular, do we have some sort of universality in its description in terms of bosonic and fermionic interacting PEPS? On the other hand, there exists several subtle issues 
in the bosonic chiral PEPS, e.g., the existence of redundant chiral branches in the ES \cite{poilblanc2015chiral,poilblanc20162,hackenbroich2018interplay,niu2022chiral} and the challenge in accessing the complete set of topological sectors \cite{chen2018non}. It is interesting to see whether fermionic PEPS describes the edge theory of CSLs faithfully, which could also shed light on resolving the problems in bosonic PEPS.
For that purpose, we study generic chiral spin liquids using optimized GfPEPS parton wavefunctions constructed from a parent Chern insulator Hamiltonian~\footnote{In Ref.~\cite{yang2015chiral} a $\mathrm{SU}(2)$ breaking chiral topological state constructed from a specific gapless topological superconductor was studied. In this paper, our scheme targets $\mathrm{SU}(2)$ symmetric chiral spin liquids constructed from generic chiral topological insulators with arbitrary Chern number.}.

\begin{figure}[t]
\centering
\includegraphics[width=\columnwidth]{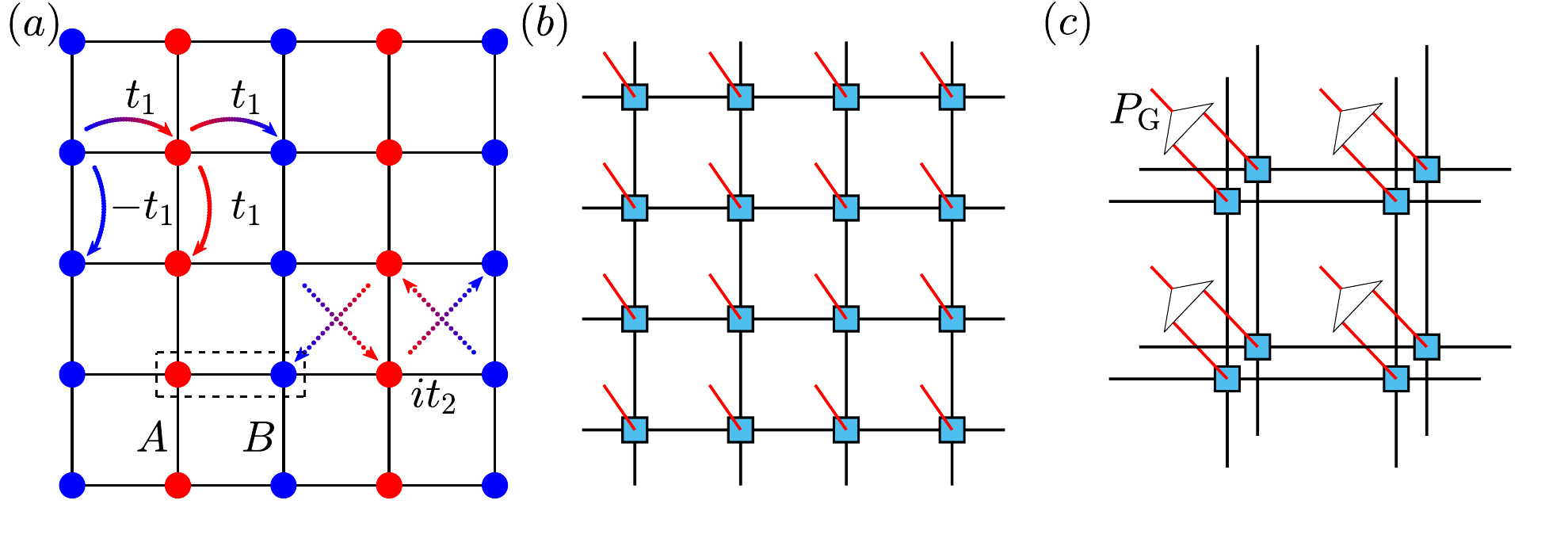}
\caption{Schematic diagrams of (a) the Hofstadter Chern insulator model with a two-site $A,B$ unit-cell along $x$ direction marked by the dashed line, (b) the translation invariant GfPEPS ansatz with $A,B$ physical sites included in one tensor and (c) the spin state constructed from Gutzwiller projected GfPEPSs.
}
\label{fig:sketch}
\end{figure}

\emph{GfPEPS for Chern insulator.}
---As a preliminary step before constructing Gutzwiller-projected parton wavefunctions, we investigate the GfPEPS representations for free fermion Chern insulators. One representative lattice model is the two-band Hofstadter model \cite{wang2013tunable,chen2021Abelian}
\begin{align}
H=&-\sum_{m,n}(t_{1}c_{m+1,n}^{\dagger}c_{m,n}+t_{1}e^{im\pi}c_{m,n+1}^{\dagger}c_{m,n}) \notag\\ 
&-\sum_{m,n}(t_{2}e^{i(m\pi\pm\pi/2)}c_{m\pm1,n+1}^{\dagger}c_{m,n})+\mathrm{H.c.}.
\label{eq:Hofstadter}
\end{align}
Here $(m,n)$ denotes the coordinates of the fermionic creation and annihilation operators, and the phases of hopping terms $t_1,t_2$ can be read from Fig.~\ref{fig:sketch} (a), providing a homogeneous $\pi/2$ flux in all triangular units. The sites can be relabeled with $A,B$ sublattice indices as $c_{2x-1,y}^{\dagger}=c_{x,y,A}^{\dagger}$ and $c_{2x,y}^{\dagger}=c_{x,y,B}^{\dagger}$.
At half-filling, the exact ground state is a gapped insulator with Chern number $C=1$ for $t_1,t_2>0$. 
To simulate the free fermion ground state, we adopt the translation invariant particle number conserving $U(1)$ symmetric GfPEPS ansatz parametrized by a single tensor in Fig.~\ref{fig:sketch} (b) and perform variational optimization \cite{mortier2022tensor,li2023u}. The translation invariant GfPEPS at half-filling can be written as a product state in the Brillouin zone, where all $k$ modes are determined by the single real space tensor and can not vary independently. The cost function is chosen as the expectation value of Eq. \eqref{eq:Hofstadter} \cite{li2023u}. Here a single tensor contains $A,B$ physical sites of the unit-cell with a physical Hilbert space dimension $d=2^2$ and virtual bond dimension $D=2^M$, where $M$ is the number of virtual fermionic modes. Thus the one-site translation (projective) symmetry in the $x$-direction is only approximately realized but can be improved with increasing $M$.

\begin{figure}[t]
\centering
\includegraphics[width=\columnwidth]{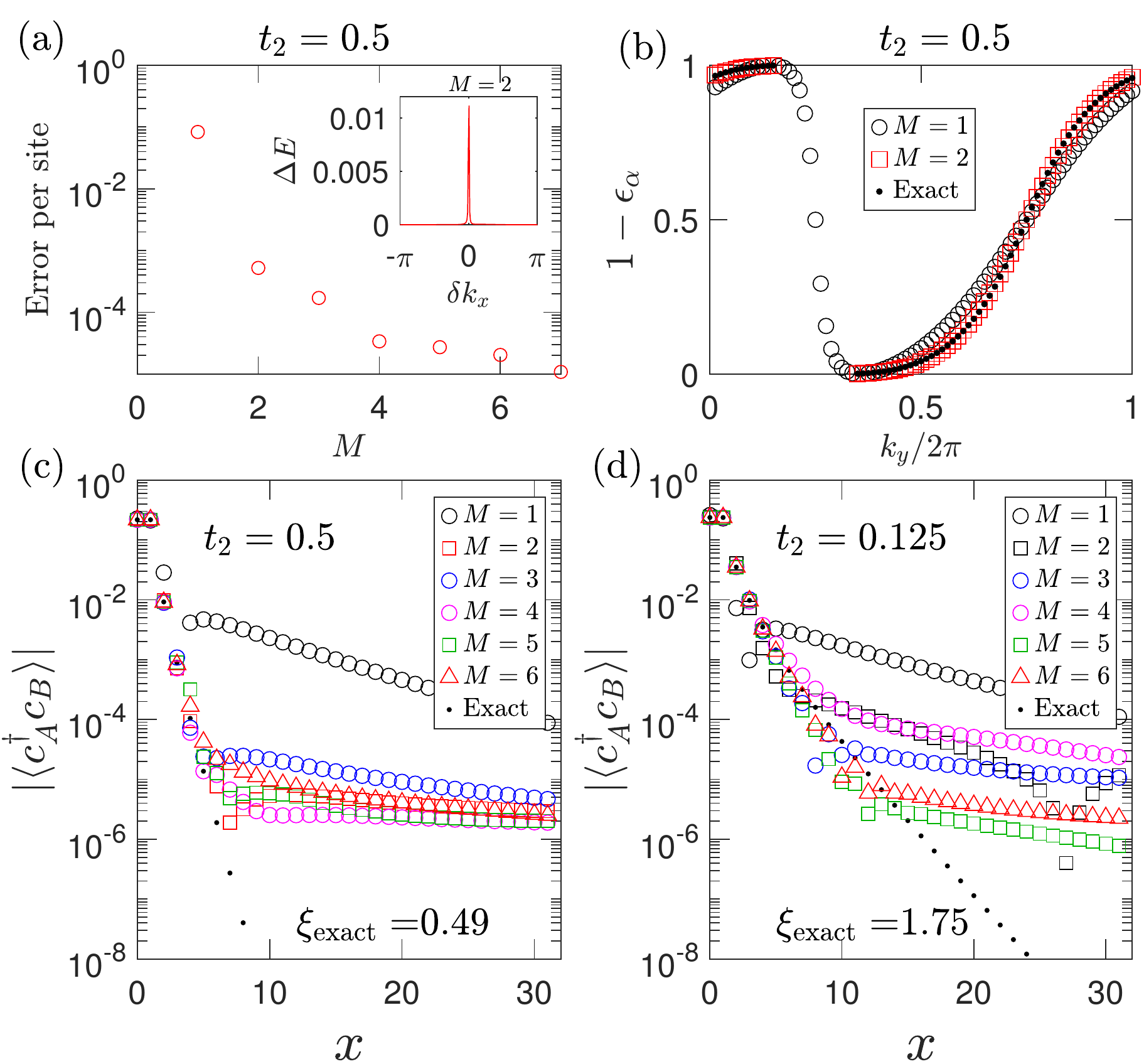}
\caption{Observables of the GfPEPS for the Hofstadter model optimized on a $80\times 80$ torus.  (a) Energy error per site versus $M$. Inset shows energy error along the $k_{x}$ direction path across the sharp singular point in $k$ space with $M=2$. (b) Edge spectrum of the correlation matrix localized at one boundary of the $40\times 80$ cylinder cut from the torus. (c)-(d) Correlation functions for different $t_2$ values. The open black circles correspond to trivial states without chirality.}
\label{fig:observables_Hofstadter}
\end{figure}

In the Hofstadter model, the optimized GfPEPS shows topological features when
the number of virtual modes satisfies $M\ge M_{\rm min}$ and then becomes sharper with increasing $M$, as depicted in Fig.~\ref{fig:observables_Hofstadter}.  We set $t_1=1$ and focus on the parameter $t_2=0.5$ with the largest band gap. Starting from $M=1$, the energy error decreases systematically, see Fig.~\ref{fig:observables_Hofstadter} (a). The topology of the optimized free fermion states can be deduced from the number of chiral branches in the single-particle ES $\lambda_{\alpha}$ \cite{peschel2003calculation,turner2010entanglement}, or, equivalently, from the edge spectrum $\epsilon_{\alpha}=(e^{\lambda_{\alpha}}+1)^{-1}$ of the subsystem correlation matrix $C^{\rm cut}$, defined as 
\begin{align}
C_{i,j}^{\text{cut}}=&\begin{cases}
\text{Tr}[|\psi\rangle\langle\psi|c_{i}^{\dagger}c_{j}], & i,j\in\text{subsystem},\\
0, & \text{otherwise}.
\end{cases}
\label{eq:C_cut}
\end{align}
Here $|\psi\rangle$ is the free fermion many-body state on the whole lattice. Along $y$ direction, we cut out a cylinder from the torus as a subsystem, and plot the correlation matrix spectrum $\epsilon_{\alpha}$ in Fig.~\ref{fig:observables_Hofstadter}(b), where bulk states have been removed according to a numerical criterion $|\epsilon_{\alpha}-0.5|>0.499$. The $M=1$ state is  non-chiral since both left-moving and right-moving modes exist, while for $M\ge 2$ the dispersion of the edge mode becomes chiral and shows quantitative agreement with the exact results.

We then examine the real space bulk correlation functions between $A,B$ sublattices at distance $x$, defined as $\langle c_{A}^{\dagger}c_{B} \rangle=\langle c_{1,y}^{\dagger} c_{2x,y}\rangle$ (the exact correlation functions between the same sublattice always vanish). The corresponding GfPEPS results are shown in Fig.~\ref{fig:observables_Hofstadter} (c) where, for $M\ge 2$, the optimized states with correct topology exhibit a crossover behaviour: at short distance the correlations decay exponentially as expected until approaching a small magnitude around $10^{-5}$, and then shows a weak long-distance gossamer tail with algebraic decay (which we have confirmed by fitting on much larger clusters). The existence of the long-distance tail can be understood from a sharp momentum space singular point as shown in the inset of Fig.~\ref{fig:observables_Hofstadter} (a) and is consistent with the topological obstruction for GfPEPS. It is expected that the correlation functions improve as $M$ increases although, in practice, the precision of our numerical optimization sets some limit. In Fig.~\ref{fig:observables_Hofstadter} (d) we show the results for $t_2=0.125$ exhibiting a much longer bulk correlation length and slower decay of correlations, from which one can roughly observe that the weight of the artificial gossamer tail decreases with $M$. 

We find a different scenario for the optimized GfPEPS in another Chern insulator model --- the Qi-Wu-Zhang model~\cite{qi2006topological}. 
The minimal bond dimension to observe the chiral edge is $M=1$ but, in that case, there is no sharp singularity in momentum space and no crossover behaviour in correlation functions, akin to the family of states investigated in Refs. \cite{wahl2013projected,wahl2014symmetries,yang2015chiral}. For larger bond dimensions $M>1$ the same momentum and real space behaviours as those in the Hofstadter model are observed. The corresponding numerical results are shown in the Supplemental Materials (SM)~\footnote{In Supplemental Materials, we supplement results for optimized GfPEPSs from other Chern insulator models, provide predicted level countings for relevant CSLs, and show ES evolution across the topological transition of the Gutzwiller projected $C=2$ parton wavefunctions.}.

\begin{figure}[t]
\centering
\includegraphics[width=\columnwidth]{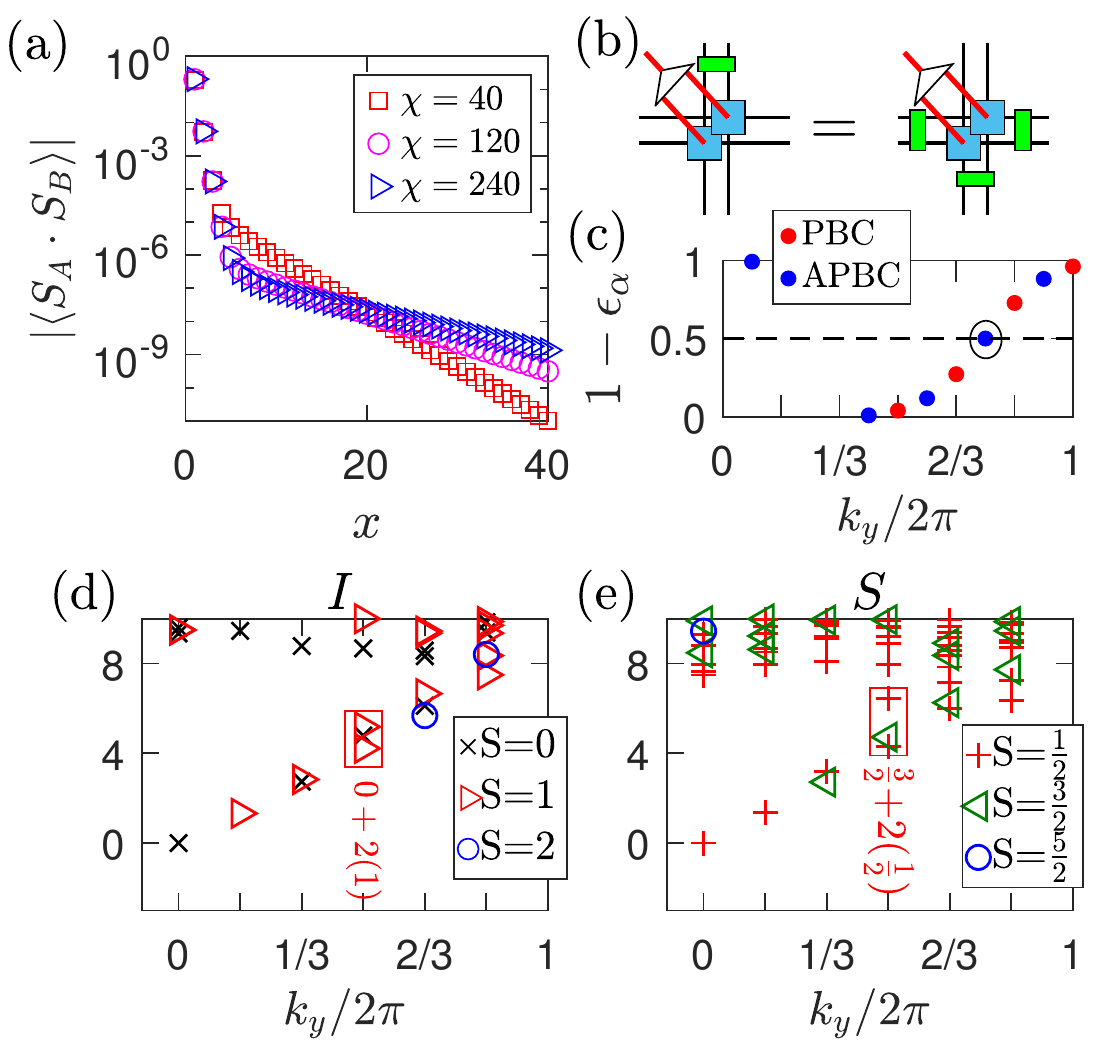}
\caption{Features of projected and unprojected $C=1$ states obtained from optimizing the Hofstadter model at $t_2=0.5$ and $M=2$. (a) Correlation functions after Gutzwiller projection, computed with various $\chi$. (b) Gauge symmetry in the local tensor of Gutzwiller projected state. (c) Edge spectrum of the free fermion correlation matrix on a width $N_y=6$ cylinder with PBC and APBC. The dashed line denotes the Fermi level of the entanglement Hamiltonian. (d)-(e) Entanglement spectra of the identity ($I$) and semion ($S$) sectors in the Gutzwiller projected states on the width $N_y=6$ cylinder with PBC and APBC, respectively. CTMRG boundary tensors with $\chi=110$ are used.}
\label{fig:ES_C1}
\end{figure}

\emph{Gutzwiller projected spin state with $C=1$.}---We now move to the Gutzwiller projected state, which is expected to be the $\mathrm{SU}(2)_1$ CSL when $C=1$. By construction \cite{yang2015chiral,li2023u}, we build the $\mathrm{SU}(2)$ invariant fermionic state via stacking two copies of GfPEPSs labeled by spin $\uparrow$ and $\downarrow$ components, hence the tensors with virtual bond dimension $4^M$ satisfy $U(1)\times \mathrm{SU}(2)$ symmetry and each virtual state is labeled by both charge and spin quantum numbers. The tensor for spin state is obtained by applying the Gutzwiller projector $P_{G}=\prod_i (\hat{n}_{i,\uparrow}+\hat{n}_{i,\downarrow})(2-\hat{n}_{i,\uparrow}-\hat{n}_{i,\downarrow})$ as shown in Fig.~\ref{fig:sketch} (c). We choose the $M=2$ GfPEPS optimized at $t_2 =0.5$, and construct PEPS representation of the projected state using the fermionc PEPS approach \cite{barthel2009contraction,corboz2010simulation}. To inspect the real space correlation functions on the infinite lattice, we use the corner transfer matrix renormalization group (CTMRG) \cite{ctmrg1,ctmrg2} method, where the approximate contraction is controlled by the environment bond dimension $\chi$, and becomes exact in the $\chi\rightarrow \infty$ limit. The numerical results for spin-spin correlations $\langle \mathbf{S}_A \cdot \mathbf{S}_B \rangle$ between $A,B$ sublattices at distance $x$ are shown in Fig.~\ref{fig:ES_C1}~(a). The correlations of the projected state decay also exponentially at short distance up to a length scale $x\approx 5$ and then decay much slower. As the absolute value of the slope at long-distance decreases with $\chi$, we expect the exact correlation function (which corresponds to the limit $\chi\rightarrow \infty$) of this $M=2$ state decays slower than any exponential decay, similar to the correlations in CSLs represented by bosonic PEPS \cite{poilblanc2017,chen2018non,hasik2022,niu2022chiral}.    

The topological order of the CSL state is characterized by the bipartite ES, which can be computed on an infinite long cylinder \cite{formalism2013}. The topologically degenerate spin states can be constructed from projected free fermion states with different boundary conditions. The flux inserted by the anti-periodic boundary condition (APBC) is realized by applying a non-contractable loop of gauge symmetry operator $Z=\prod_i Z_i$ on the virtual space \cite{toriccode2009}, where the gauge symmetry operator $Z_i$ takes the form $Z_i=(-1)^{n_i}$, as illustrated in Fig.~\ref{fig:ES_C1}(b). Here the gauge symmetry can be interpreted as the fermion parity of virtual spin-$1/2$ particles or the number parity of singlet pairs crossing the $i$th virtual bond~\cite{schuch2012,poilblanc2012topological}. The $\mathrm{SU}(2)_1$ CSLs on the cylinder have two topological sectors. On a finite cylinder, the minimally entangled states (MES) \cite{zhang2012quasiparticle,tu2013momentum} are determined by explicitly controlling populations of edge modes in the unprojected states. Correspondingly, on the infinite cylinder we determine the MES according to the single-particle ES in the unprojected states as well as virtual space quantum numbers, 
based on the equivalence between edge spectrum and entanglement spectrum as implied by Eq.~\eqref{eq:C_cut} \cite{peschel2003calculation,turner2010entanglement}.

In order to control the filling of the free fermion edge modes, we plot in Fig.~\ref{fig:ES_C1}(c) the spectrum 
of the subsystem correlation matrix representing a (single) physical edge. The Fermi level $\epsilon_{F}=0.5$ ($\lambda_{F}=0$ marked by a dashed line) defines the Fermi sea state $|\psi_{\rm FS}\rangle=\prod_{\sigma=\uparrow,\downarrow}\prod_{1-\epsilon_{\alpha}<\epsilon_{F}}d_{\alpha,\sigma}^{\dagger}|\rm Vac\rangle$, where $d_{\alpha,\sigma}^{\dagger}$ denote the bulk and edge modes with eigenvalue $\epsilon_{\alpha}$ and spin polarization $\sigma$. The $\mathrm{SU}(2)_1$ ground states in identity ($I$) and semion ($S$) sectors can be constructed on the PBC/APBC cylinder as \cite{wu2020tensor}

\begin{align}
|\psi_{I}\rangle &=P_G|\psi_{\rm FS}\rangle_{\rm PBC},\notag \\ 
|\psi_{S}\rangle_{\sigma,\bar{\sigma}} &=P_G\zeta_{L,\sigma}^{\dagger}\zeta_{R,\bar{\sigma}}^{\dagger}|\psi_{\rm FS}\rangle_{\rm APBC},
\end{align}
respectively. Here $\zeta_{L/R,\sigma}^{\dagger}$ creates the lowest particle excitation at the left/right boundary as marked by the black circle, and the superposition $|\psi_{S}\rangle_{\uparrow,\downarrow}-|\psi_{S}\rangle_{\downarrow,\uparrow}$ forms a singlet.
The two topological sectors have 
a total spin difference $\Delta S=1/2$ in the horizontal virtual space, and can be distinguished by the $y$-direction loop operator
$P_{\rm even/odd} =(1\pm Z)/2$
that projects to the subspaces of integer spin (even charge parity) and half-integer spin (odd charge parity) in the ES, respectively.

Fig.~\ref{fig:ES_C1}(d)-(e) shows numerical results of the Gutzwiller projected state on the $N_{y}=6$ cylinder for $I$ and $S$ sectors obtained with PBC and APBC respectively, where the entanglement Hamiltonians \cite{cirac2011entanglement} are built from CTMRG boundary tensors \cite{poilblanc20162,chen2018non,chen2020}. Numerically the integer (half-integer) sector corresponds to the fixed point of the transfer matrix with PBC (APBC), in agreement with the fact that for both cases the unprojected edge modes are half-filled (Fig.~\ref{fig:ES_C1} (c)). The low energy levels match the prediction of $\mathrm{SU}(2)_1$ WZW CFT (see SM), including the ones marked by red rectangles. We notice that compared to the previous bosonic PEPS method \cite{poilblanc2015chiral,poilblanc20162}, our fermionic construction yields the correct number of chiral branches in the ES. We also remark that we expect both sectors can be obtained within a fixed boundary condition as long as $N_y$ is large enough with sufficient number of linear edge states as in Fig.~\ref{fig:observables_Hofstadter}(b).

\emph{Gutzwiller projected spin states with $C=2$.}---
Above approach can be naturally generalized to parton states with arbitrary Chern number. Here we consider a $C=2$ model~\cite{zhang2014identifying},
which turns out to be
nontrivial as it shows that the topological order of a generic projected parton state depends not only on the Chern number before projection, but also on details of the wavefunctions.

\begin{figure}[t]
\centering
\includegraphics[width=\columnwidth]{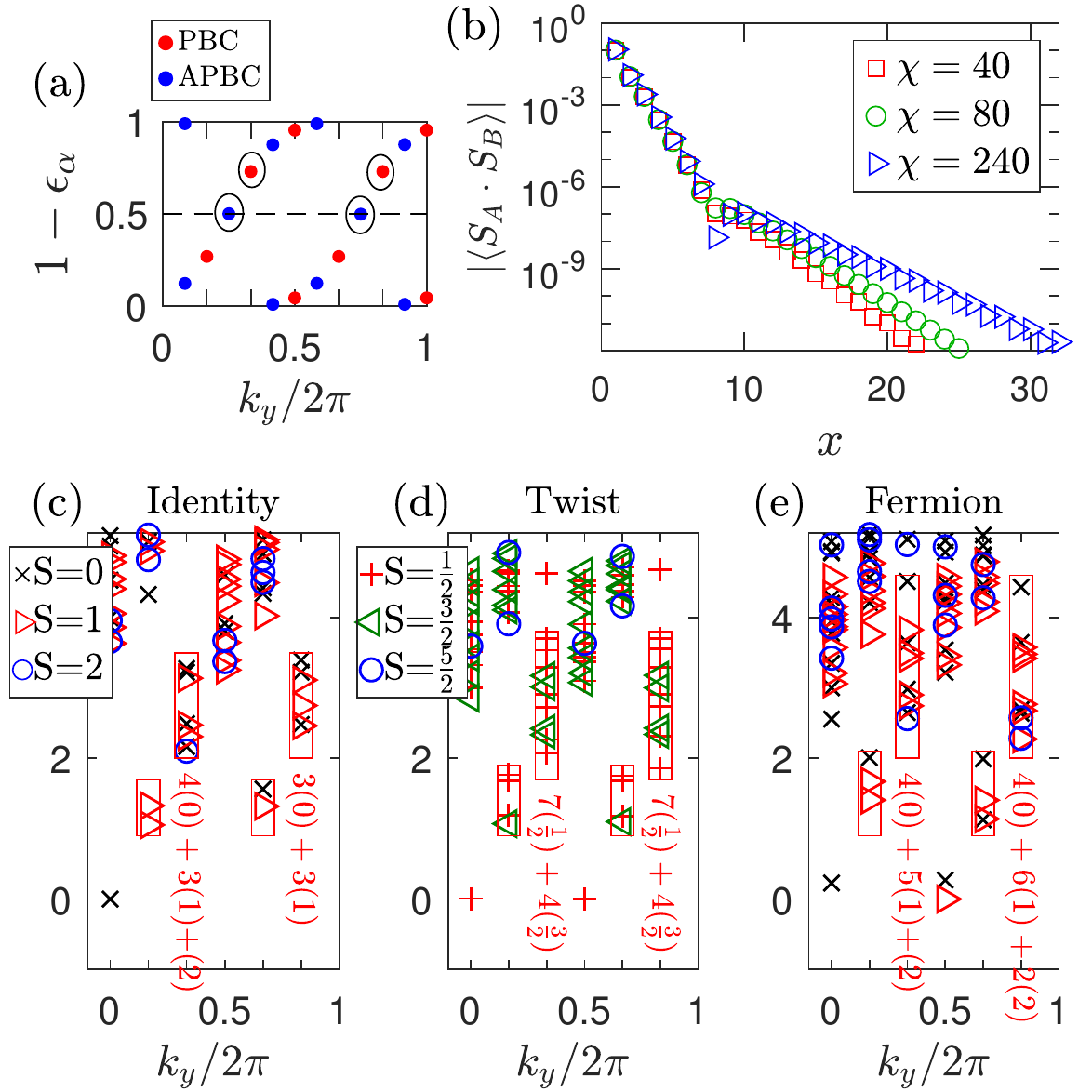}
\caption{Features of the unprojected and projected $C=2$ states with $M=2$ at $\Theta=\pi/4$. (a) Edge spectrum of the free fermion correlation matrix on a width $N_y=6$ cylinder with PBC and APBC. (b) Correlation functions after Gutzwiller projection computed at different $\chi$.  (c)-(e) The ES of $\mathrm{SO}(5)_1$ CSL at $\Theta=\pi/4$. In (c) and (e) PBC is used while in (d) APBC is used. For (c)-(e) CTMRG boundary tensors with $\chi=110$ are used.}
\label{fig:ES_C2}
\end{figure}

The family of free fermion $C=2$ Hamiltonians $H_{\Theta}$ have $A,B$ sublattices in the unit-cell. At $\Theta=0$ it takes the form 
\begin{align}
H_{\Theta=0}=&\sum_{\langle i,j \rangle_{x}}t_{1}(c_{j,A}^{\dagger}c_{i,B}+c_{j,B}^{\dagger}c_{i,A}) \notag\\ 
+&\sum_{\langle i,j \rangle_{y}}t_{1}(c_{j,A}^{\dagger}c_{i,A}-c_{j,B}^{\dagger}c_{i,B}) \notag\\ 
+&\sum_{\langle\langle i,k \rangle\rangle}t_{2}e^{2i\theta_{ik}}(c_{k,B}^{\dagger}c_{i,A}-c_{k,A}^{\dagger}c_{i,B})+\mathrm{H.c.},
\label{eq:Hofstadter_C2}
\end{align}
where $i,j,k$ denotes the sites on the $x-y$ plane and $\theta_{ik}$ denotes the angle between next nearest neighbour sites $i,k$. The model at $\Theta=0$ can be viewed as two independent layers of Eq.~\eqref{eq:Hofstadter} that differs by a one-site translation $T_x$ along $x$ direction: by taking $A$ sites for even $x$ coordinate and taking $B$ sites for odd $x$ coordinate one obtains the first copy, and vice versa for the second copy. Due to the $T_x$ translation the entanglement spectrum along $y$ direction cut contributed from two layers are identical but has $\pi$ momentum difference, as shown in Fig.~\ref{fig:ES_C2} (a) for optimized $M=2$ GfPEPS with PBC and APBC (with a string along $x$ direction inserted), respectively. Applying the local unitary $U(\Theta)=\exp[\sum_{i} \Theta(c_{i,A}^{\dagger}c_{i,B}-c_{i,B}^{\dagger}c_{i,A})/2]$ that acts inside each unit-cell, the family of Hamiltonian $H_{\Theta}=U^{-1}(\Theta)H_{0}U(\Theta)$ is obtained. At $\Theta=0$ the two $C=1$ layers are independent and at $\Theta=\pi/4$ the two layers are maximally mixed. The total Chern number and free fermion entanglement spectrum do not depend on $\Theta$ since $U(\Theta)$ is local. 

After Gutzwiller projection, a topological transition emerges along the path $\Theta\in[0,\pi/4]$ (see SM). At $\Theta < \Theta_c$ the projected state is the Abelian $\mathrm{SU}(2)_1 \times \mathrm{SU}(2)_1$ since the $\Theta=0$ gapped topological phase of the two decoupled layers should have a finite extension in parameter space. We focus on the $\mathrm{SO}(5)_1$ CSL realized around the maximally mixed limit $\Theta=\pi/4$ which was predicted by the effective field theory and verified recently by a matrix product state calculation~\cite{zhang2014identifying,wu2022non}. Before investigating the topological properties of this non-Abelian CSL we emphasize that correlations of the projected GfPEPS (Fig.~\ref{fig:ES_C2}(b)) show a similar crossover behavior as in Fig.~\ref{fig:ES_C1}(a), pointing towards the universality of such artifact in chiral PEPS.

In the free fermion edge spectrum on the left boundary of the $N_y=6$ cylinder, we denote $\zeta_{L,1,\sigma}^{\dagger},\zeta_{L,2,\sigma}^{\dagger}$ as the first excited states with momentum difference $\pi$ marked by the black circles in Fig.~\ref{fig:ES_C2}(a) for both PBC and APBC.   
For the $\mathrm{SO}(5)_1$ CSL, three MESs~\cite{wu2022non} can be constructed as 
\begin{align}
|\psi_{\rm Identity}\rangle &=P_{G}|\psi_{\rm FS}\rangle_{\rm PBC},\notag \\
|\psi_{\rm Twist}\rangle_{\sigma,\bar{\sigma}} &=P_{G}\zeta_{L,a,\sigma}^{\dagger}\zeta_{R,a,\sigma}|\psi_{\rm FS}\rangle_{\rm PBC},\notag \\
|\psi_{\rm Fermion}\rangle &=P_{G}\zeta_{L,a,\uparrow}^{\dagger}\zeta_{L,a,\downarrow}^{\dagger}\zeta_{R,b,\uparrow}^{\dagger}\zeta_{R,b,\downarrow}^{\dagger}|\psi_{\rm FS}\rangle_{\rm APBC}.
\label{eq:MES_SO5}
\end{align}
Here $a,b\in\{1,2\}$. For the twist sector, due to the annihilation operator $\zeta_{R,a,\sigma}$ a single spin $\bar{\sigma}$ edge mode is left at the right boundary.
The numerical results for the projected states at $\Theta=\pi/4$ are shown in Fig.~\ref{fig:ES_C2} (c)-(e). With PBC and APBC, the dominant integer spin sectors which have half-filled edge modes are shown to be the identity and fermion sectors, respectively. The half-integer (odd charge parity) sector with PBC gives the twist sector. The $\Delta k_y=\pi$ momentum splitting of the chiral branches originates from the existence of two edge branches shifted by $\pi$ before projection. A key advantage of the fermionic approach is that all topological sectors can be explicitly constructed, which is not obvious to achieve within the bosonic PEPS framework~\cite{chen2018non}.
The level counting of the numerically obtained ES shows a remarkable agreement with the CFT prediction, 
which is given in the SM.


\emph{Conclusion.}---We have investigated the Gutzwiller projected Chern insulators in the GfPEPS representation. The topological obstruction for chiral GfPEPS only leads to very weak Gossamer tails in the correlation functions of projected GfPEPSs and thus brings no practical obstruction for numerical simulations. Within our framework, topological sectors can be tuned conveniently by flux insertion without explicit control of edge mode populations, and the projected GfPEPSs provide faithful descriptions of the edge spectra in both Abelian and non-Abelian CSLs, which would be interesting to further analyze via a recently proposed generalized Gibbs ensemble approach~\cite{Arildsen2022}. 
In the future, it would also be interesting to use such families of projected GfPEPS as variational manifolds to attack quantum spin models. Note that, the non-Abelian SO$(5)_1$ CSL does not seem to have a simple description with bosonic PEPS. One natural question emerges: for spin systems, is there a difference in the representative power of bosonic PEPS and fermion PEPS? We leave this question to future research.

\emph{Acknowledgement.---}We thank Hong-Hao Tu for insightful discussions. We implement non-Abelian symmetries using the TensorKit.jl package~\cite{tensorkit}. This work was granted access to the HPC resources of CALMIP center under the allocation 2017-P1231. J.-Y.C. acknowledges support by Open Research Fund Program of the State Key Laboratory of Low-Dimensional Quantum Physics (project No.~KF202207), Fundamental Research Funds for the Central Universities, Sun Yat-sen University (project No.~23qnpy60), a startup fund from Sun Yat-sen University (No.~74130-12230034), and the Innovation Program for Quantum Science and Technology 2021ZD0302100. This work was also supported by the TNTOP ANR-18-CE30-0026-01 grant awarded by the French Research Council.

\bibliography{GfPEPS.bib}

\noindent

\renewcommand{\thesection}{S-\arabic{section}}
\setcounter{section}{0}  
\renewcommand{\theequation}{S\arabic{equation}}
\setcounter{equation}{0}  
\renewcommand{\thefigure}{S\arabic{figure}}
\setcounter{figure}{0}  

\indent

\end{document}


\newcommand{\dagga}{{\phantom{\dagger}}}
\newcommand{\didier}[1]{{\color{blue}[DP: #1]}}
\newcommand{\SN}[1]{{\color{red}[SN: #1]}}

\renewcommand{\theequation}{S\arabic{equation}}
\renewcommand{\thefigure}{S\arabic{figure}}
\renewcommand{\thetable}{S\Roman{table}} 

\renewcommand*{\citenumfont}[1]{S#1}
\renewcommand*{\bibnumfmt}[1]{[S#1]}

\title{Supplemental materials}
\maketitle

\section{Optimized GfPEPS for other Chern insulator models }
As the Hofstadter model with $C=1$ has been shown in the main text, here we supplement results for the Qi-Wu-Zhang model \cite{QWZ} with $C=1$ and the model with $C=2$ \cite{Hofstadter_C2} to show that the features of the optimized chiral GfPEPS are universal.

\subsection{The Qi-Wu-Zhang model with $C=1$}
\begin{figure}[htb]
\centering
\includegraphics[width=0.5\columnwidth]{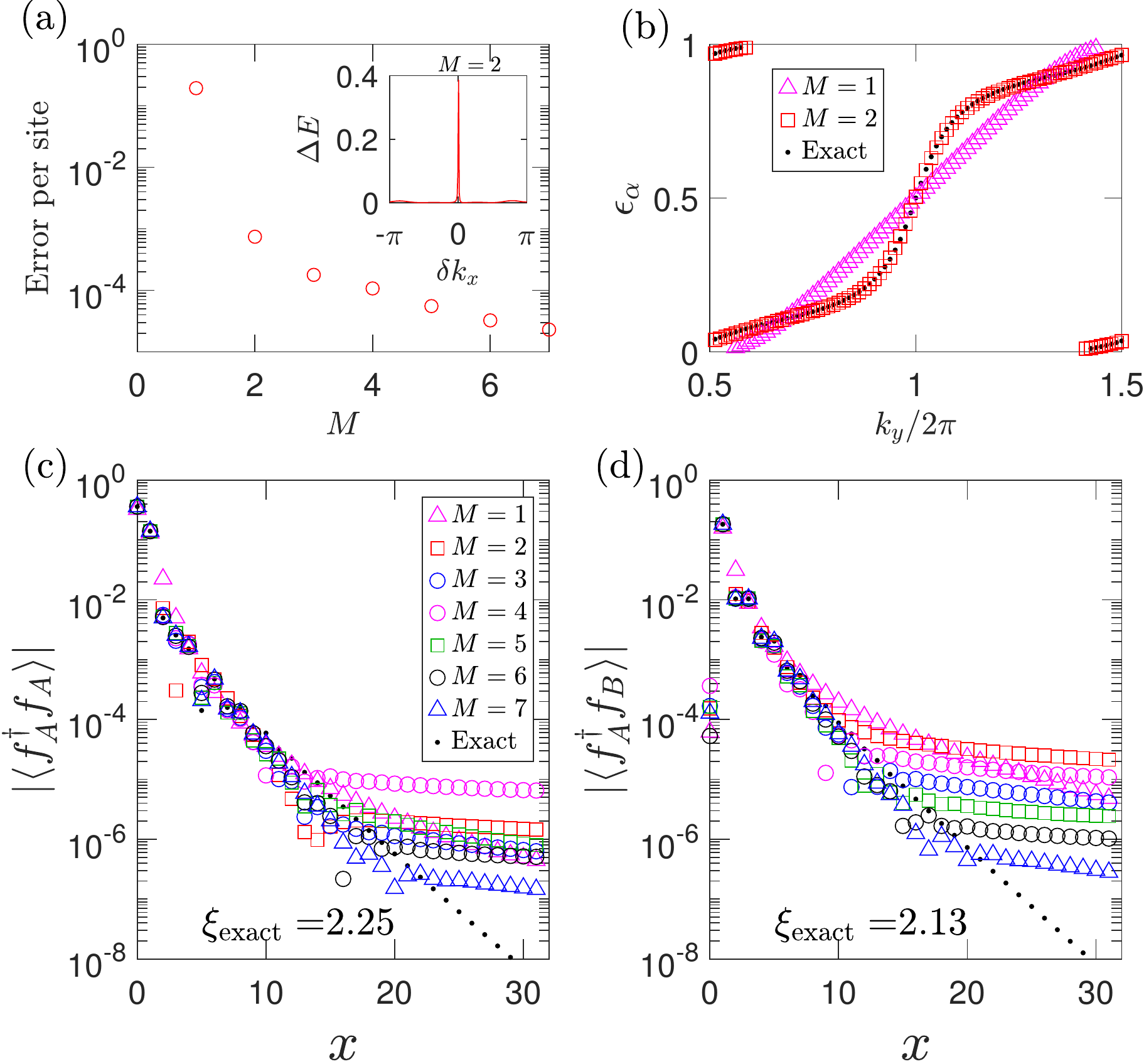}
\caption{Observables of the GfPEPS for the Qi-Wu-Zhang model optimized on a $80\times 80$ torus.  (a) Error of energy per site versus $M$. Inset shows energy error along the $k_{x}$ direction path across the sharp singular point in $k$ space with $M=2$. (b) Edge spectrum of the correlation matrix localized at one boundary of the $40\times 80$ cylinder cut from the torus. (c)-(d), Correlation functions between $AA$ and $AB$ sublattices. The correlation functions for $M=1$ GfPEPS show algebraic decay at all length scale.}
\label{fig:observables_QWZ}
\end{figure}

\begin{figure*}[htb]
\centering
\includegraphics[width=1\textwidth]{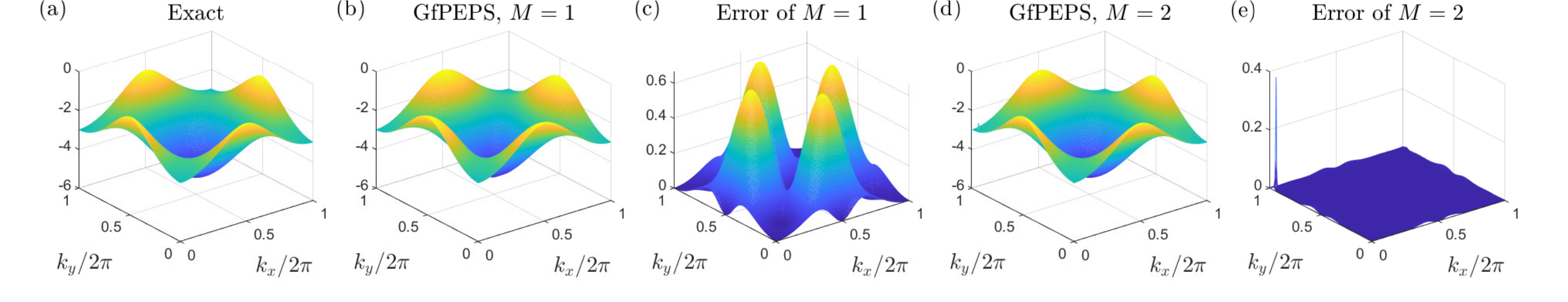}
\caption{Momentum space energy bands of optimized GfPEPS in the Qi-Wu-Zhang model. (a) The exact lower energy band. (b) The energy of $M=1$ GfPEPS evaluated at each $k$ point in Brillouin zone. (c) Energy error of the $M=1$ GfPEPS in the Brillouin zone. (d) Energy of the $M=2$ GfPEPS evaluated at each $k$ point in the Brillouin zone. (e) Energy error of the $M=2$ GfPEPS in the Brillouin zone. }
\label{fig:bands_QWZ}
\end{figure*}

The tight-binding Hamiltonian for the Qi-Wu-Zhang model in momentum space has the form
\begin{align}
H(k)=2\sin k_{y} \sigma_x+2\sin k_{x}\sigma_y+(m-2\cos k_{x}-2\cos k_{y})\sigma_z.
\end{align}
Here the Pauli matrices act on the sublattice degree of freedom and we take $m=1$. In our GfPEPS ansatz, each tensor contains two physical sites labeled by sublattice index $A,B$. 
In Fig.~\ref{fig:observables_QWZ} (a)-(d), we show the observables of optimized GfPEPS. In contrast to the Hofstadter model, the optimized $M=1$ GfPEPS for the Qi-Wu-Zhang model falls in a special family of chiral GfPEPS investigated in Refs. \cite{chiralGfPEPS1,chiralGfPEPS2,chiralGfPEPS3}. As can be seen in Fig.~\ref{fig:observables_QWZ} (b), the $M=1$ GfPEPS is chiral but the dispersion of the entanglement spectrum does not accurately reproduce the exact result. The real space correlations of the $M=1$ GfPEPS have no crossover behaviour and show algebraic decay at all length scales as can be seen in Fig.~\ref{fig:observables_QWZ} (c)-(d). Correspondingly, in momentum space there is no sharp singular point as shown in Fig.~\ref{fig:bands_QWZ} (a)-(c). For larger bond dimensions $M\ge 2$, the optimized chiral GfPEPS are similar to those optimized from the Hofstadter model, i.e. the dispersion of the entanglement spectrum becomes close to the exact behavior, the correlation functions become very accurate at short distance, and sharp singular points appear in momentum space (see Fig.~\ref{fig:bands_QWZ} (d)-(e)). 

\begin{figure*}[htb]
\centering
\includegraphics[width=0.7\textwidth]{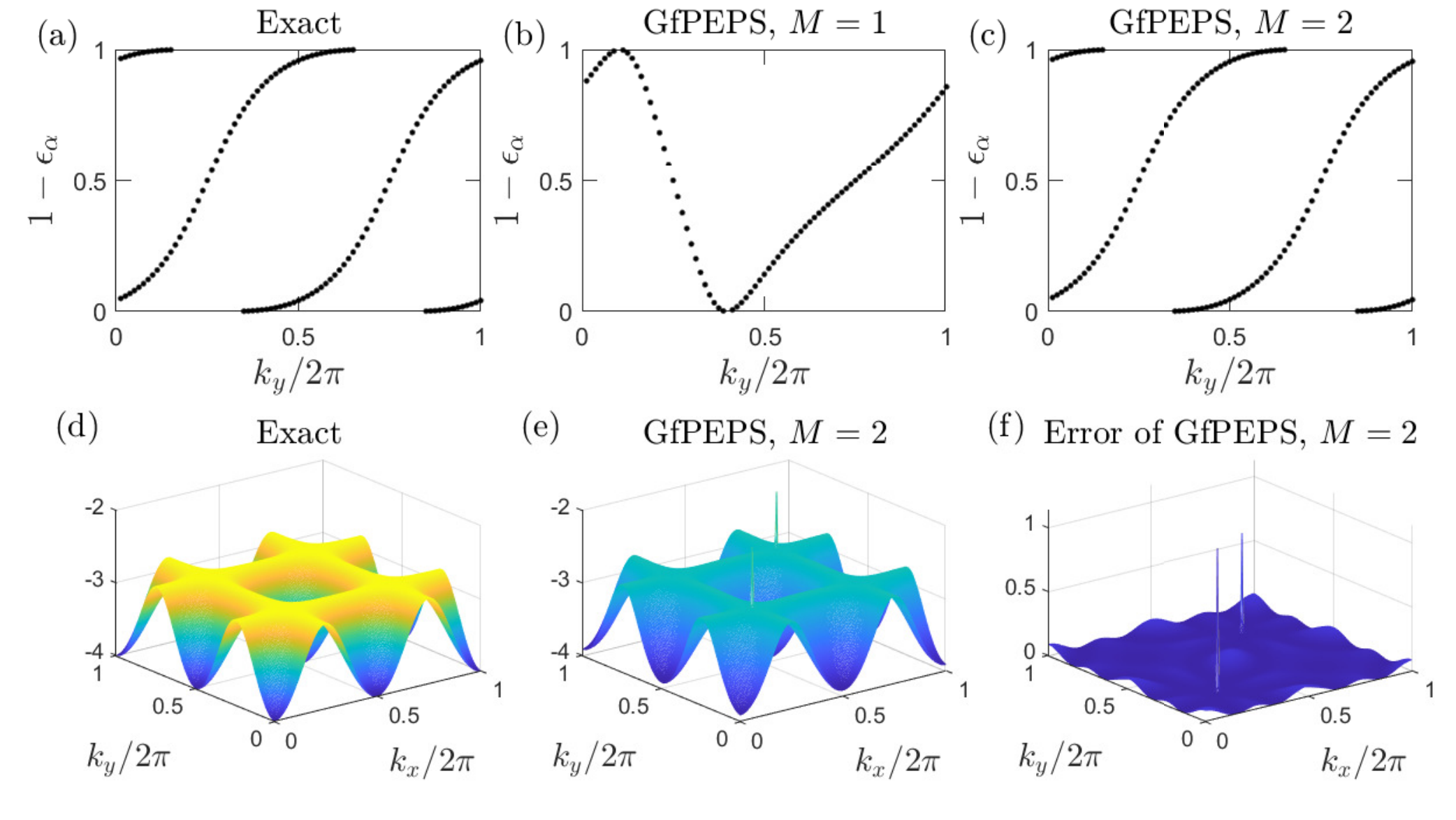}
\caption{Observables of the GfPEPS for the $C=2$ model optimized on a $80\times 80$ torus. (a)-(c) Edge spectrum of the correlation matrix localized at one boundary of the $40\times 80$ cylinder cut from the torus. (d) The exact lower energy band. (e) The energy of $M=2$ GfPEPS evaluated at each $k$ point in Brillouin zone. (f) Energy error of the $M=2$ GfPEPS in Brillouin zone. }
\label{fig:bands_C2}
\end{figure*}

\subsection{The $C=2$ model}
We also provide results of the optimized GfPEPS for the $C=2$ model (at $\Theta=0$):
\begin{align}
H_{\Theta=0}=&\sum_{\langle i,j \rangle_{x}}t_{1}(c_{j,A}^{\dagger}c_{i,B}+c_{j,B}^{\dagger}c_{i,A}) +\sum_{\langle i,j \rangle_{y}}t_{1}(c_{j,A}^{\dagger}c_{i,A}-c_{j,B}^{\dagger}c_{i,B}) \notag\\ 
+&\sum_{\langle\langle i,k \rangle\rangle}t_{2}e^{2i\theta_{ik}}(c_{k,B}^{\dagger}c_{i,A}-c_{k,A}^{\dagger}c_{i,B})+H.c..
\label{eq:Hofstadter_C2_SM}
\end{align}
The optimization does not depend on $\Theta$ since $U(\Theta)$ is a local unitary. In our GfPEPS ansatz, each tensor still contains two physical sites labeled by sublattice index $A,B$. The minimal bond dimension with $C=2$ is $M=2$, as shown in Fig.~\ref{fig:bands_C2} (a)-(c): the $M=1$ state does not have chiral modes while the $M=2$ state has two chiral branches with very accurate dispersion. The $M=2$ state has two singular points in the energy distribution in momentum space as illustrated in Fig.~\ref{fig:bands_C2} (d)-(e).

Here we make a few remarks. First, the topological obstruction for GfPEPS only applies in the thermodynamic limit, i.e., both $N_x,N_y\rightarrow \infty$. For a thin cylinder with small $N_y$ and infinite $N_x$, the correlation functions decays exponentially at long distance, as the singular $k$ points in momentum space can be avoided by discrete $k$ values. Second, as already mentioned in the main text, the topological nature of the parton wavefunction depends not only on the Chern number before projection, but also on details of the free fermion wavefunction. In this respect, the GfPEPS with larger $M$ has better quality since they represent the fermionic state more accurately.

\section{Level counting of different CSLs}

Here we show the lowest three Virasoro levels of the $\mathrm{SU}(2)_1$, $\mathrm{SU}(2)_1\times \mathrm{SU}(2)_1$ and $\mathrm{SO}(5)_1$ CSLs in Table~\ref{table:CSLs}. The results for $\mathrm{SU}(2)_1\times \mathrm{SU}(2)_1$ is computed from considering the tensor product of two branches of $\mathrm{SU}(2)_1$ CSL levels, with both the energy and the momentum of the two branches summed up. The results for the $\mathrm{SO}(5)_1$ CSL is taken from Ref.~\cite{SO5CSL}, where the $\mathrm{SO}(5)$ representations have been decomposed into $\mathrm{SU}(2)$ representations.

\begin{table}[htb]
\centering
{\renewcommand{\arraystretch}{1.5}
\renewcommand{\tabcolsep}{0.2cm}
\begin{tabular}{|c|c|c|c|c|}
\hline 
CSL & Sector & 1st & 2nd & 3rd\tabularnewline
\hline 
\multirow{2}{*}{$\mathrm{SU}(2)_{1}$} & Identity ($I$) & $0$ & $1$ & $0\oplus1$\tabularnewline
\cline{2-5} \cline{3-5} \cline{4-5} \cline{5-5} 
 & Semion ($S$) & $\frac{1}{2}$ & $\frac{1}{2}$ & $\frac{1}{2}\oplus\frac{3}{2}$\tabularnewline
\hline 
\multirow{4}{*}{$\mathrm{SU}(2)_{1}\times \mathrm{SU}(2)_{1}$} & $I\otimes I$ & $0$ & $2(1)$ & $3(0)\oplus3(1)\oplus2$\tabularnewline
\cline{2-5} \cline{3-5} \cline{4-5} \cline{5-5} 
 & $I\otimes S$ & $\frac{1}{2}$ & $2(\frac{1}{2})\oplus\frac{3}{2}$ & $4(\frac{1}{2})\oplus3(\frac{3}{2})$\tabularnewline
\cline{2-5} \cline{3-5} \cline{4-5} \cline{5-5} 
 & $S\otimes I$ & $\frac{1}{2}$ & $2(\frac{1}{2})\oplus\frac{3}{2}$ & $4(\frac{1}{2})\oplus3(\frac{3}{2})$\tabularnewline
\cline{2-5} \cline{3-5} \cline{4-5} \cline{5-5} 
 & $S\otimes S$ & $0\oplus1$ & $2(0)\oplus2(1)$ & $3(0)\oplus5(1)\oplus2(2)$\tabularnewline
\hline 
\multirow{3}{*}{$\mathrm{SO}(5)_{1}$} & Identity & $0$ & $0\oplus3(1)$ & $7(0)\oplus6(1)\oplus2$\tabularnewline
\cline{2-5} \cline{3-5} \cline{4-5} \cline{5-5} 
 & Twist & $2(\frac{1}{2})$ & $6(\frac{1}{2})\oplus2(\frac{3}{2})$ & $14(\frac{1}{2})\oplus8(\frac{3}{2})$\tabularnewline
\cline{2-5} \cline{3-5} \cline{4-5} \cline{5-5} 
 & Fermion & $2(0)\oplus1$ & $3(0)\oplus4(1)$ & $8(0)\oplus11(1)\oplus3(2)$\tabularnewline
\hline 
\end{tabular}}
\caption{\label{table:CSLs} Topological sectors and level countings predicted from the Wess–Zumino–Witten CFT for different CSLs.}
\end{table}
\section{Topological transition of the projected $C=2$ states}
\begin{figure*}[htb]
\includegraphics[width=1\textwidth]{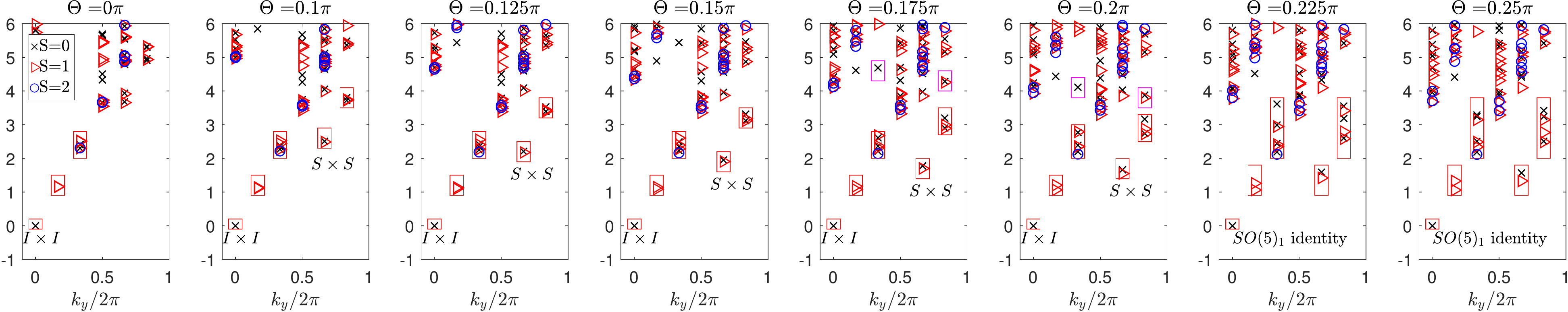}
\caption{\label{fig:Sector1}
Evolution of the entanglement spectrum from the $I\times I$ sector of the $\mathrm{SU}(2)_1 \times \mathrm{SU}(2)_1$ CSL to the identity sector of the $\mathrm{SO}(5)_1$ CSL on the $N_y=6$ cylinder. The pink boxes mark the levels that are absorbed during the transition. PBC and projector $P_{\rm even}$ are used. Entanglement Hamiltonians are constructed from CTMRG tensors with $\chi=110$.}
\end{figure*}

\begin{figure*}[htb]
\includegraphics[width=1\textwidth]{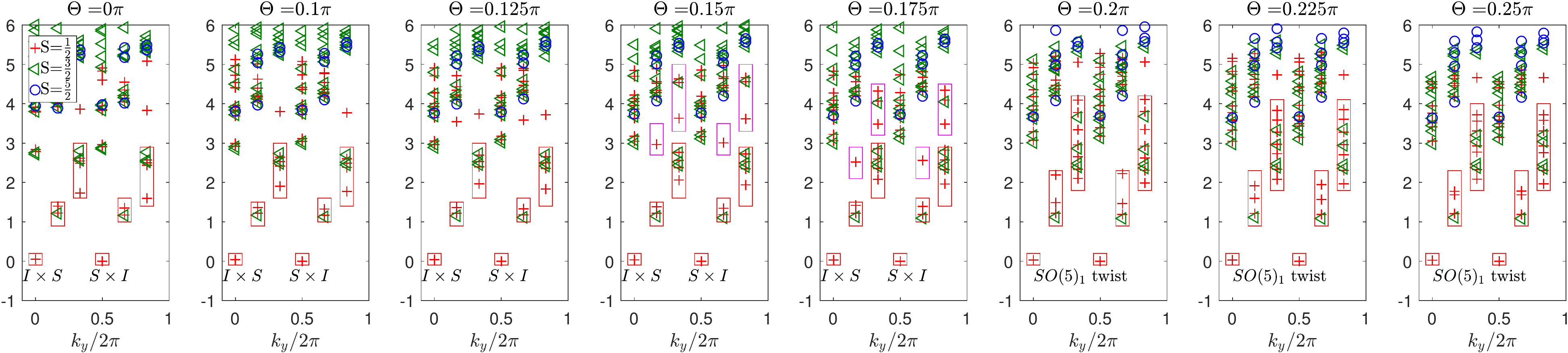}
\caption{\label{fig:Sector2}
Evolution of the entanglement spectrum from the $I\times S \oplus S\times I$ sector of the $\mathrm{SU}(2)_1 \times \mathrm{SU}(2)_1$ CSL to the twist sector of the $\mathrm{SO}(5)_1$ CSL on the $N_y=6$ cylinder. The pink boxes mark the levels that are absorbed during the transition. PBC and projector $P_{\rm odd}$ are used. Entanglement Hamiltonians are constructed from CTMRG tensors with $\chi=110$.}
\end{figure*}

\begin{figure*}[htb]
\includegraphics[width=1\textwidth]{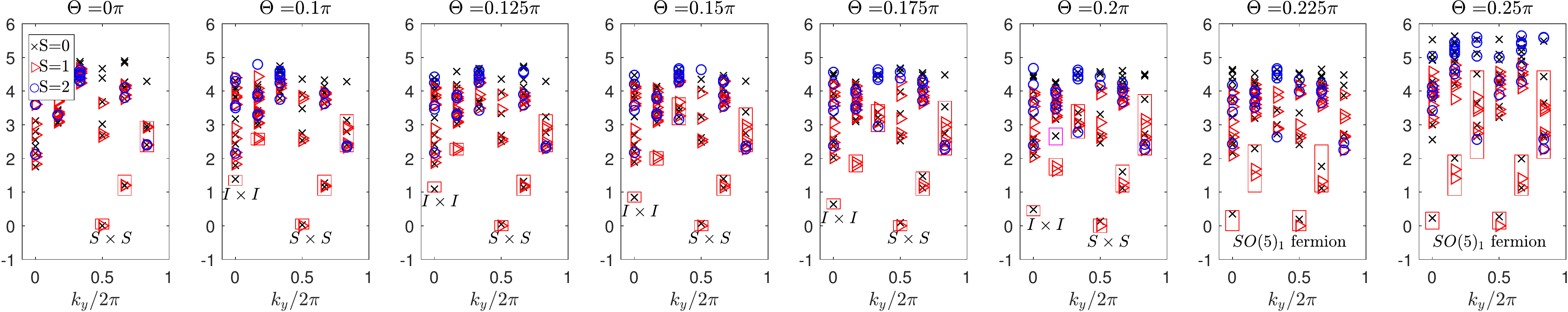}
\caption{\label{fig:Sector3}
Evolution of the entanglement spectrum from the $S\times S$ sector of the $\mathrm{SU}(2)_1 \times \mathrm{SU}(2)_1$ CSL to the fermion sector of the $\mathrm{SO}(5)_1$ CSL on the $N_y=6$ cylinder. The pink boxes mark the levels that are absorbed during the transition. APBC and projector $P_{\rm even}$ are used. Entanglement Hamiltonians are constructed from CTMRG tensors with $\chi=110$.}
\end{figure*}

The Hamiltonian $H_{\Theta}$ along the path is defined by $H_{\Theta}=U^{-1}(\Theta)H_{0}U(\Theta)$. Since $U(\Theta)=\exp[\sum_{i} \Theta(c_{i,A}^{\dagger}c_{i,B}-c_{i,B}^{\dagger}c_{i,A})/2]$ acts locally on each tensor, we just need to optimize GfPEPS at $\Theta=0$ and apply the local unitary to obtain the GfPEPS tensor for any $\Theta$. We compute the ES along the path $\Theta\in[0,\pi/4]$ on a $N_y=6$ cylinder with fixed boundary conditions as displayed in Fig.~\ref{fig:Sector1}, \ref{fig:Sector2} and \ref{fig:Sector3}. The topological order on the $\Theta$ path can be identified according to the level counting in Table~\ref{table:CSLs}. The critical value $\Theta_c$ for the topological transition on our $N_y=6$ cylinder (which is the largest $N_y$ accessible to us) is $\Theta_c \approx 0.2\pi$, in agreement with the finite size effect observed in the VMC study~\cite{Hofstadter_C2} i.e. $\Theta_c$ decreasing with increasing system size. 
Below we analyze the evolution of the ES along the path.  

In the limit $\Theta\rightarrow 0$, the global state is a tensor product of two $\mathrm{SU}(2)_1$ CSLs, the MESs can be inferred from the knowledge of the $\mathrm{SU}(2)_1$ CSL. The $2\times 2=4$ MESs can be constructed as
\begin{align}
|\psi_{I\times I}\rangle &=P_{G}|\psi_{\rm FS}\rangle_{\rm PBC},\notag \\
|\psi_{S\times S}\rangle_{\sigma,\bar{\sigma}}^{\sigma',\bar{\sigma'}}
&=P_{G}\zeta_{L,1,\sigma}^{\dagger}\zeta_{L,2,\sigma'}^{\dagger}\zeta_{R,1,\bar{\sigma}}^{\dagger}\zeta_{R,2,\bar{\sigma}'}^{\dagger}|\psi_{\rm FS}\rangle_{\rm APBC},\notag \\
|\psi_{S\times I}\rangle^{\sigma,\bar{\sigma}} &=P_{G}\zeta_{L,1,\sigma}^{\dagger}\zeta_{R,1,\sigma}|\psi_{\rm FS}\rangle_{\rm PBC},\notag \\
|\psi_{I\times S}\rangle_{\sigma,\bar{\sigma}} &=P_{G}\zeta_{L,2,\sigma}^{\dagger}\zeta_{R,2,\sigma}|\psi_{\rm FS}\rangle_{\rm PBC}.
\end{align}
With PBC the modes $\zeta_{R,1,\sigma},\zeta_{R,2,\sigma}$ denote the lowest hole excitation on the right boundary, hence a single spin $\bar{\sigma}$ edge mode is left at the right boundary of $I\times S$ or $S\times I$ sector. 
As can be inferred from the results of the $\mathrm{SU}(2)_1$ CSL, $I\times I$ sector (with PBC) and $S\times S$ sector (with APBC) are realized with half-filled edge modes and correspond to the largest fixed points in the integer spin sector. Hence for PBC and APBC the integer spin sectors are shown to be the $I\times I$ and $S\times S$ sectors respectively, see Fig.~\ref{fig:Sector1} and \ref{fig:Sector2}. The half-integer spin fixed point with PBC turns out to be the superposition of $I\times S$ and $S\times I$, see Fig.~\ref{fig:Sector3}.

For generic $\Theta$ in the $\mathrm{SU}(2)_1\times \mathrm{SU}(2)_1$ phase, the integer spin fixed point with both PBC and APBC is a superposition of $I\times I$ and $S\times S$ sectors [Fig.~\ref{fig:Sector1}, \ref{fig:Sector2}], as the two $\mathrm{SU}(2)_1$ CSLs are coupled and the two MESs can not be distinguished by quantum numbers. Nevertheless, we can still identify the superposition of MESs from the level counting, as marked by black labels at the bottom of the spectra. 
Interestingly, the evolution of the ES across the transition, as shown in Fig.~\ref{fig:Sector1}-\ref{fig:Sector3}, tells us how the topological sectors of the unfamiliar non-Abelian $\mathrm{SO}(5)_1$ CSL is related to the well-known $\mathrm{SU}(2)_1$ CSL: For PBC and integer spin, the $I\times I$ sector in $\mathrm{SU}(2)_1\times \mathrm{SU}(2)_1$ superposes with the high energy $S\times S$ sector and then forms the $\mathrm{SO}(5)_1$ identity sector. For PBC and half-integer spin, the superposition of the $I\times S$ and $S\times I$ sectors forms the $\mathrm{SO}(5)_1$ twist sector by absorbing high energy levels marked by pink rectangles. For APBC and integer spin, the $S\times S$ sector superposes with the $I\times I$ sector at higher energy and then forms the $\mathrm{SO}(5)_1$ fermion sector. 

